\documentclass{PoS}

\title{INTEGRAL observes the 2007 outburst of the Be transient SAX~J2103.5+4545}

\ShortTitle{INTEGRAL observes the 2007 outburst of the Be transient SAX~J2103.5+4545}

\author{\speaker{L.~Ducci}$^{ab}$,
  L.~Sidoli$^b$, A.~Paizis$^b$, S.~Mereghetti$^b$,
  P.~M.~Pizzochero$^c$\\
  \llap{$^a$} Dipartimento di Fisica e Matematica, Universit\`a degli
  Studi dell'Insubria,\\ 
  Via Valleggio 11, I-22100 Como, Italy \\
  \llap{$^b$} INAF, Istituto di Astrofisica Spaziale e Fisica
  Cosmica,\\
  Via E. Bassini 15, I-20133 Milano, Italy \\
  \llap{$^c$} Dipartimento di Fisica, Universit\`a degli Studi di
  Milano, \\
  Via Celoria 16, I-20133 Milano, Italy \\

  E-mail: \email{lorenzo@iasf-milano.inaf.it}  
}

\abstract{We performed a detailed study of the 2007 outburst of the
  $352 \ s$ pulsar SAX~J2103.5+4545, a Be/X-ray transient observed by
  \emph{INTEGRAL}, to study  its spectral and temporal properties
  during the evolution of the outburst. 
  SAX~J2103.5+4545 was observed with IBIS/ISGRI from 25 to
  27 April 2007 and from 6 to 8 May 2007. The $20-100 \ keV$ spectrum
  is well described by a bremsstrahlung model with a temperature $kT
  \sim 24 \ keV$. The pulse profiles are variable with time and
  energy. A pulse period derivative of 
  $\dot{P}_{spin-up}=(-3.4 \pm 0.4)\times 10^{-7} \ s \ s^{-1}$ has
  been observed during the outburst. Instead, a spin-down of 
  $\dot{P}_{spin-down}=(5.5 \pm 0.4)\times 10^{-9} \ s \ s^{-1}$ is
  observed between the 2007 outburst reported here and the previous
  one occurred in December 2004. This is the largest spin-down
  measured for SAX~J2103.5+4545 since its discovery. We estimate a
  neutron star magnetic field in the range $(1.6-3)\times
  10^{13} \ G$ using the Ghosh \& Lamb torque model.}

\FullConference{7th INTEGRAL Workshop\\
		 September 8-11 2008\\
		 Copenhagen, Denmark}

\begin{document}

\section{Introduction}

SAX J2103.5+4545 is a transient Be/X-ray binary
with an orbital period of $12.67$~days \cite{Camero Arranz et al. 2007}
hosting a pulsar with 
a pulse period of 352.2~s \cite{Baykal et al. 2007}.
It was discovered during an outburst
in February 1997 \cite{Hulleman et al. 1998}, and since then
numerous type I and type II outbursts have been observed 
\cite{Camero Arranz et al. 2007}.
The optical counterpart is a B0Ve star with highly variable $H\alpha$ emission, 
located at a distance of $6.5 \pm 0.9$~kpc \cite{Reig et al. 2004}.
The position of SAX~J2103.5+4545 in the 
spin period versus orbital period diagram deviates
from the correlation of HMXB with Be star and falls in the region
typical of wind-fed HMXB with OB supergiant companions.
The X-ray energy spectrum in the 5--200~keV band is
described by a power-law with exponential cut-off, with photon index 
$\Gamma = 1.53$, cut-off energy $E_c = 19$~keV
and e-folding energy $E_F = 32$~keV.
No cyclotron lines have been observed \cite{Sidoli et al. 2005}.
A soft blackbody component with temperature kT~$\sim$~1.9~keV 
was observed with \emph{XMM-Newton} \cite{Inam et al. 2004},
and an iron emission line at $\sim$~6.42~keV 
was observed with \emph{RXTE} \cite{Baykal et al. 2002}.
SAX~J2103.5+4545 has shown numerous spin-up phases during its outbursts.
A correlation between X-ray flux and spin-up rate has been found
\cite{Baykal et al. 2000} which can be explained with  
the formation of an accretion disk.

\section{Observations and data analysis}

SAX~J2103.5+4545 underwent an outburst in April--May 2007
\cite{Galis et al. 2007}.
This outburst was observed by the IBIS/ISGRI instrument 
on 25--27 April 2007 ($54215.39-54217.02$ MJD)
and on 6--8 May 2007 ($54226.71-54228.98$ MJD).

\begin{figure}
  \centering
  \includegraphics[width=8cm]{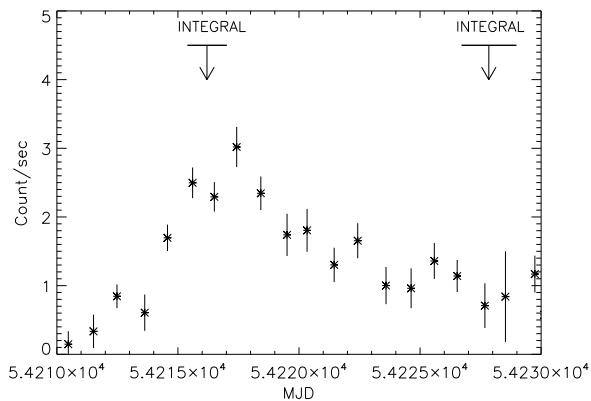}
  \caption{\emph{RXTE}/ASM light-curve of SAX J2103.5+4545 
           in the 2-10~keV energy range           
           (54210 -- 54230 MJD, April 20 -- May 10 2007).
           The arrows indicate the time interval covered 
           by the \emph{INTEGRAL} observations.}
  \label{SAXJ2103_ASM_2007_2}
\end{figure}

Here we report the data analysis 
of the source field observed with IBIS-ISGRI.
The \emph{Off-line Scientific Analysis} (OSA 6.0)
software was used. 
We have divided this observations 
in 2 groups (Table \ref{INTEGRAL_observations})
which are marked in Figure \ref{SAXJ2103_ASM_2007_2},
where the 2--10~keV light-curve 
of the 2007 outburst  
observed by \emph{RXTE}/ASM is reported.

\begin{table}
\begin{center}
\begin{tabular}{lcccc}
\hline
\hline
Obs. &  Time interval    &   Start Time   &   End Time   &   Exposure time \\
     &                   &     (MJD)      &    (MJD)     &       (ks)      \\
\hline
1    & 25-27 April 2007  &   $54215.39$   &  $54217.02$  &      $129$      \\
2    &   6-8 May   2007  &   $54226.71$   &  $54228.98$  &      $169$      \\
\hline
\end{tabular}
\end{center}
\caption{The two groups of IBIS/ISGRI observations of SAX~J2103.5+4545 used
  in this work.}
\label{INTEGRAL_observations}
\end{table}

We extracted two spectra of SAX~J2103.5+4545 
in the energy range $22-100$~keV,
in the time intervals 25--27 April 2007 and 6--8 May 2007
(Table \ref{INTEGRAL_observations}).
A systematic uncertainty of 1\% was added quadratically
to the statistical errors to account for uncertainties in the 
instrumental response.

We performed timing analysis in the range $15-60$~keV,
where the best signal to noise ratio is obtained.
We corrected the arrival times to the Solar System barycenter
and for Doppler delay due to the pulsar orbital motion.
The correction of the arrival times can be written as:
\begin{equation} \label{delta_t}
\delta t = x \sin l -\frac{3}{2}xe \sin \omega  +\frac{1}{2}xe\cos \omega \sin (2l) -\frac{1}{2} xe \sin \omega \cos (2l)
\end{equation}
where $x=a_x \sin(i)/c$ is the light--travel time for the projected
semi-major axis $a_x$, $i$ is the inclination angle,
$e$ is the eccentricity and $\omega$
is the longitude of periastron \cite{Deeter et al. 1981}.
$l$ is the mean orbital longitude at time $t$ which is given by:
\begin{equation} \label{longitudine}
l = 2 \pi \frac{(t-T_{\pi/2})}{P_{orb}} +\frac{\pi}{2}
\end{equation}
where $T_{\pi/2}$ is the epoch when the mean orbital longitude 
is equal to $\pi/2$ and $P_{orb}$ is the orbital period.
To correct the arrival times, 
we have assumed the most recent orbital parameters
obtained by Camero Arranz et al. (2007) \cite{Camero Arranz et al. 2007}.
An epoch folding technique was used to measure 
the pulse period \cite{Leahy 1987}.

\section{Results}

\subsection{Spectral analysis}

\begin{figure*}
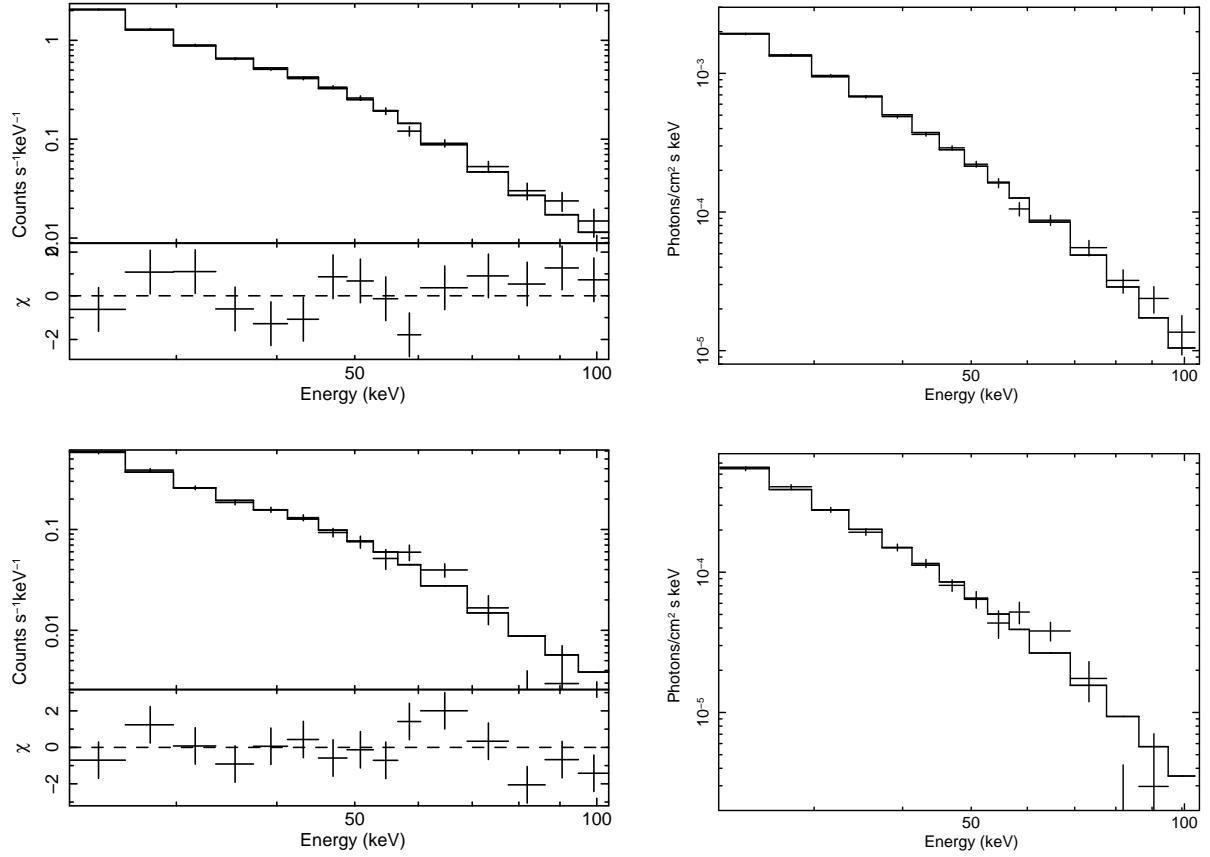

\centering
\begin{tabular}{cc}
\includegraphics[height=8.3cm, angle=-90]{./ps_files/NUOVO_spettro_SAX_553_22_100keV_bremss.ps} & \includegraphics[height=7.4cm, angle=-90]{./ps_files/NUOVO_spettro_SAX_553_22_100keV_ufs.ps} \\

\includegraphics[height=8.3cm, angle=-90]{./ps_files/NUOVO_spettro_SAX_0557_22_100keV_bremss.ps} & \includegraphics[height=7.4cm, angle=-90]{./ps_files/NUOVO_spettro_SAX_557_22_100keV_ufs.ps}
\end{tabular}
\caption{ISGRI SAX~J2103.5+4545 spectra in the energy range 22--100~keV, 
         on 25-27 April 2007 (\emph{Top}) and on 6--8 May 2007 (\emph{Bottom}).
         \emph{Left panels:} counts spectra, best--fit bremsstrahlung spectra and residuals
         (in units of standard deviations);
         \emph{Right panels:} photon spectra together with the best--fit model.}
\label{spettro_553}
\end{figure*}

We have fitted the $22-100$~keV spectra with several simple models
and found that the best $\chi^2$ value is given by a bremsstrahlung model
with a temperature $kT \approx 24$~keV for both spectra 
(see Figure \ref{spettro_553} for energy spectra, 
and Table \ref{tabella_spettri_553} for parameters).

The average flux (20--100~keV) obtained from the spectrum 
in the time interval 25--27 April 2007 
is $F_x = (1.68{+0.09 \atop -0.07}) \times 10^{-9}$~erg~cm$^{-2}$~s$^{-1}$,
while on 6--8 May 2007 
the flux is $F_x = (5.0{+0.06 \atop -0.06}) \times 10^{-10}$~erg~cm$^{-2}$~s$^{-1}$,
translating into a luminosity $L_x = (8.5 \pm 2.5)\times 10^{36}$~erg~s$^{-1}$
(25--27 April 2007) and  $L_x = (2.5 \pm 0.7) \times 10^{36}$~erg~s$^{-1}$
(6--8 May 2007) for a distance $d=6.5\pm0.9$~kpc. 
These luminosities are similar
to those measured during previous outbursts
(\cite{Sidoli et al. 2005}, \cite{Baykal et al. 2007}).

   \begin{table}
\begin{center}
\begin{tabular}{lcccc}
\hline
\hline
Time interval     &             kT          &  $\chi^2_{\nu}$   &      Flux (20 - 40 keV)          &       Flux (20 - 100 keV)       \\
                  &            (keV)        &  for 13 (d.o.f.) &  $10^{-9}$~erg~cm$^{-2}$~s$^{-1}$  &   $10^{-9}$~erg~cm$^{-2}$~s$^{-1}$  \\
\hline
25--27 April 2007 & $23.4{+0.8 \atop -0.7}$ &      1.06        &     $1.11 {+0.04 \atop -0.03}$    &     $1.68 {+0.09 \atop -0.07}$   \\
\\
6--8 May 2007     & $24 {+2 \atop -2}$      &      1.29        &          $0.32 \pm 0.03$          &           $0.50 \pm 0.06$        \\
\hline
\end{tabular}
\end{center}
      \caption{Best fit parameters of the two average spectra of 
               SAX~J2103.5+4545 with a bremsstrahlung model.}
         \label{tabella_spettri_553}
\end{table}

\subsection{Timing analysis}

We measured the pulse period and the associated uncertainty
for each of the two data sets reported in Table \ref{INTEGRAL_observations}
by means of an epoch folding technique, obtaining the values 
reported in Table \ref{tabella_periodi_pulsazione}.
\begin{table}
\begin{center}
\begin{tabular}{lc @{ \ \ \ \ \ } c}
\hline
\hline
Name &  Time interval    & $P_{spin}$ (s)       \\
\hline
$P_1$    & 25-27 April 2007  & $352.725 \pm 0.004$ \\
$P_2$    &   6-8 May   2007  & $352.382 \pm 0.004$ \\
\hline
\end{tabular}
\end{center}
\caption{Pulse periods of SAX~J2103.5+4545.}
\label{tabella_periodi_pulsazione}
\end{table}
The average spin-up rate between $P_1$ and $P_2$ is
$ \dot{P}_{spin-up} = (-3.4 \pm 0.4) \times 10^{-7} \ \mbox{s s}^{-1} \label{Ppunto spin-up} $
The average flux during this spin-up is 
$ \bar{F}_x = 3.6 \times 10^{-9} \ erg \ cm^{-2}\ s^{-1} $
in the energy range $1-200$~keV.

\begin{figure}
\centering
\includegraphics[width=8.5cm, angle=-90]{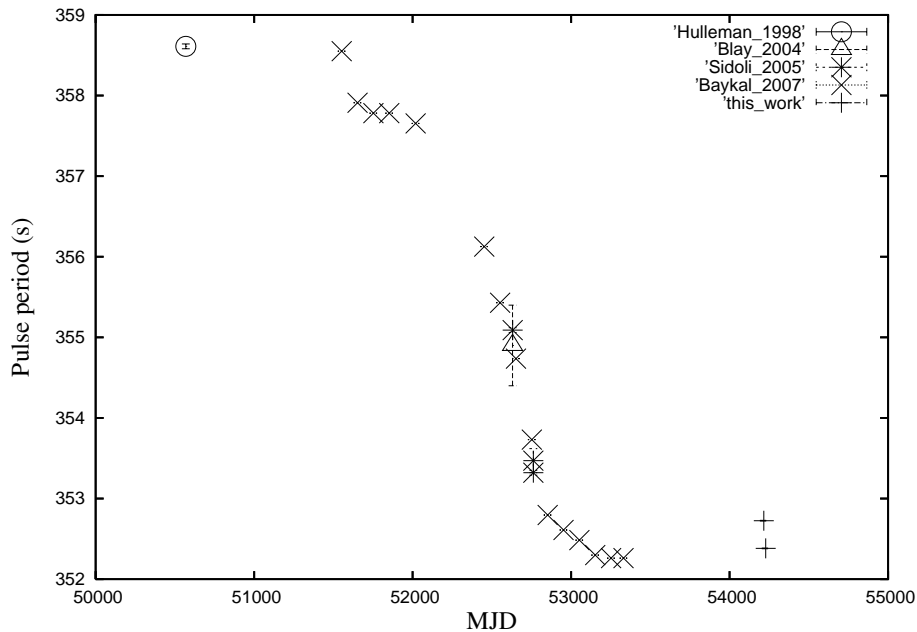}
\caption{Pulse period history of SAX~J2103.5+4545. The pulse period
  values are taken by \emph{BeppoSAX} \cite{Hulleman et al. 1998},
  \emph{INTEGRAL} \cite{Blay et al. 2004}, \cite{Sidoli et al. 2005}, 
  and \emph{RXTE} \cite{Baykal et al. 2007}.}
\label{evol_puls_dal1997}
\end{figure}

SAX~J2103.5+4545 has shown a pulse period decrease since 1997,
occasionally interrupted by spin-down intervals,
corresponding to low accretion rates.
(\cite{Sidoli et al. 2005}, \cite{Baykal et al. 2007}).
In particular, the pulse period history 
(Figure \ref{evol_puls_dal1997}) 
shows a spin-down between December 2004 ($53340 \pm 10$~MJD) 
where $P_{Baykal \ 2007} = 352.31 \pm 0.01$~s \cite{Baykal et al. 2007}, 
and the last outburst (April - May 2007, 
see Table \ref{INTEGRAL_observations}), 
when we measured $P_{1} = 352.725 \pm 0.004$~s.
Between these two outbursts SAX~J2103.5+4545 was in a low luminosity state.
We measured the upper limit flux of SAX~J2103.5+4545
during this low luminosity state from December 2004 to April 2007
using \emph{RXTE}/ASM data.
We found $F < 2 \times 10^{-11} \ erg \ cm^{-2}\  s^{-1} $
in the energy range $2-10$~keV.
The average pulse period derivative from December 2004 to April 2007
is $ \dot{P}_{spin-down}=(5.5 \pm 0.4) \times 10^{-9} \ s \ s^{-1} $.

\begin{figure}
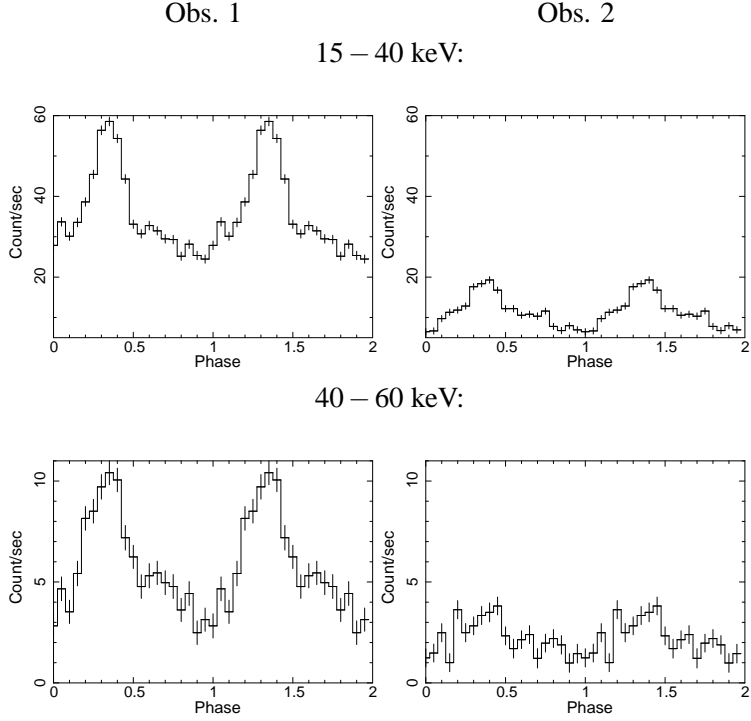

\centering
\begin{tabular}{@{}c@{}c@{}}
\ \ Obs. 1  & \ \ Obs. 2 \\
\multicolumn{2}{@{}c@{}}{\ \ $15-40$~keV:} \\
\includegraphics[width=3.7cm, angle=-90]{./ps_files/efold_SAXJ2103_riv553_15_40keV_51000MJD.ps} & \includegraphics[width=3.7cm, angle=-90]{./ps_files/efold_SAXJ2103_riv557_15_40keV_51000MJD.ps} \\
\multicolumn{2}{@{}c@{}}{\ \ $40-60$~keV:} \\
\includegraphics[width=3.7cm, angle=-90]{./ps_files/efold_SAXJ2103_riv553_40_60keV_51000MJD.ps} & \includegraphics[width=3.7cm, angle=-90]{./ps_files/efold_SAXJ2103_riv557_40_60keV_51000MJD.ps} 
\end{tabular}
\caption{Evolution of SAX~J2103.5+4545 pulse profiles for different time intervals 
  and for two energy bands. From top to bottom: $15-40 \ keV$, $40-60 \ keV$.
  From left to right: Obs. 1, Obs. 2 (see Table 3).}
\label{forma_profili_pulsazione_dipendente_dall_energia}
\end{figure}

For each of the 2 data sets reported in Table \ref{INTEGRAL_observations}, 
we have obtained the pulse profiles 
for the energy bands 15--60~keV, 15--40~keV and 40--60~keV.
The pulse profiles are binned into 20 phase bins
(zero phase at 51000~MJD; see Figure \ref{forma_profili_pulsazione_dipendente_dall_energia}).
We have found a temporal variability
for the pulse profiles, as already observed during
previous outbursts of this source 
(\cite{Camero Arranz et al. 2007}, \cite{Sidoli et al. 2005}).

\section{Discussion and Conclusions}

SAX J2103.5+4545 is a Be/X-ray binary with transient
X-ray emission, fed by the wind of the Be companion.
In previous outbursts a correlation between X-ray flux 
and spin-up rate was found 
which can be explained with the formation of an accretion disk
\cite{Baykal et al. 2000}.

When a neutron star is fed by an accretion disk,
the material which flows towards the neutron star magnetic poles
produces X-ray luminosity and at the same time 
a torque on the rotating neutron stars
mediated by its magnetosphere.
The angular momentum of the infalling matter
is transferred to the pulsar at the magnetospheric boundary
\cite{Ghosh and Lamb 1979}.
The Ghosh and Lamb model assumes a magnetically-threaded disk
allowing not only spin-up, but also spin-down,
which occurs when the magnetic and corotation radii are comparable
and the luminosity is low.
Ghosh and Lamb predicted the following correlation
between the pulse period
derivative and the X-ray luminosity:
\begin{equation} \label{equa_ghoshlamb}
\dot{P} = -5.0 \times 10^{-5} \mu_{30}^{2/7} n(\omega_s) R_6^{6/7} I^{-1}_{45} (P L_{37}^{3/7} )^2 \left ( \frac{M_{NS}}{M_{\odot}}\right )^{-3/7} \ \ s \ yr^{-1} 
\end{equation}
where $\mu_{30}$ is the neutron star magnetic dipole moment  
in the disk plane ($\mu = BR^3$) in units of $10^{30} \ G \ cm^3$,
$R_6$ is the stellar radius in units of $10^6 \ cm$, $M_{NS}$ 
is the mass of the neutron star in grams,
$I_{45}$ is the moment of inertia in units of $10^{45} \ g \ cm^2$, 
$P$ is the pulse period, $L_{37}$ is the accretion luminosity
in units of $10^{37} erg \ s^{-1}$.
We have considered the dimensionless torque 
obtained by Ghosh \& Lamb (1979) \cite{Ghosh and Lamb 1979}:
\begin{equation} \label{torque Ghosh e Lamb}
n(\omega_s) \approx 1.39\frac{1 - \omega_s[4.03(1-\omega_s)^{0.173} - 0.878]}{1 - \omega_s}
\end{equation}
where $\omega_s$ is called \emph{fastness parameter}
and is defined as:
\begin{equation} \label{fastness parameter}
\omega_s \equiv \frac{\Omega_s}{\Omega_K(r_0)} = \left ( \frac{r_0}{R_c} \right )^{3/2} 
\end{equation}
where $\Omega_s$ is the angular velocity of the star, 
$\Omega_K(r_0) = (GM_{NS}/r_0^3)^{1/2}$ is the Keplerian
angular velocity at $r_0$, the radius which divides the outer part
of the accretion disk, where the velocity is keplerian,
from the inner part, where the plasma corotates with the star
\cite{Ghosh and Lamb 1979};
$R_c = (GM_{NS}/\Omega_s^2)^{1/3}$ is the corotation radius of the star.
Therefore, during an outburst, a variation in the accretion rate
can modify the pulse period derivative and at the same time 
the X-ray luminosity.
In this framework it is possible to
determine the magnetic field of SAX~J2103.5+4545.
\begin{figure}[htbp]
\begin{center}
  \includegraphics[width=10.5cm]{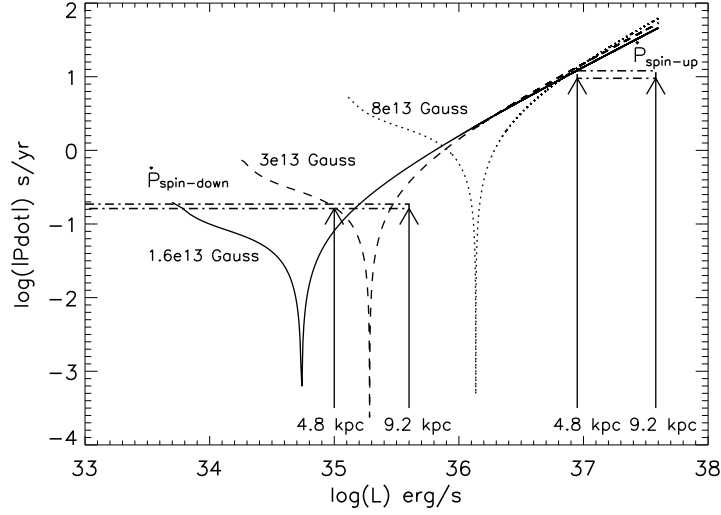}
\end{center}
\caption{Comparison between the observed spin-up and spin-down rate and the theoretical relation 
         between $\dot{P}$ and the source luminosity. 
         The box labeled with $\dot{P}_{spin-up}$ indicates the spin-up rate
         measured during the time interval 25 April - 8 May 2007,
         the box labeled with $\dot{P}_{spin-down}$ indicates the spin-down rate
         measured during the time interval December 2004 - April 2007.
         The fit with theoretical curves is obtained 
         for a distance $\sim 4.8$~kpc
         and a magnetic field $B=1.6 \times 10^{13}$~G (solid line) 
         and $B=3 \times 10^{13}$~G (dashed line).
         The uncertainties of $\dot{P}_{spin-up}$, $\dot{P}_{spin-down}$ and $L_x$ 
         are given by the pulse period derivative uncertainties
         and by the uncertainty of the source distance.
         }
\label{grafico_ghosh_lamb}
\end{figure}
Using equations
(\ref{equa_ghoshlamb}) to (\ref{fastness parameter}),
we assumed $M_{NS} = 1.4 \ M_{\odot}$, $R_{NS} = 10$~km,
an average pulse period $\bar{P}=352.5$~s,
$\bar{F}_x = 3.6 \times 10^{-9}$~erg~cm$^{-2}$~s$^{-1}$
as average flux during the spin-up,
and the upper limit flux during the spin-down. 
We found a magnetic field of $(1.6 - 3)\times 10^{13} \ G$
for a distance of $\sim 4.8 \ kpc$.
This distance is consistent with what found by Reig et al. (2004),
if the error in the absolute magnitude found 
by Vacca et al. (1996) \cite{Vacca et al. 1996} is included.
These results are also consistent with what previously obtained by
Baykal et al. (2007) \cite{Baykal et al. 2007}.
In Figure (\ref{grafico_ghosh_lamb}) we compare the spin-up and
spin-down rate observed with \emph{INTEGRAL} with the theoretical
relation between $\dot{P}$ and the source luminosity for 3 different
magnetic fields of the neutron star. Vertical lines mark the source
luminosities for $d=4.8 \ kpc$ and $d=9.2 \ kpc$.

During the low luminosity state (December 2004 - April 2007)
the upper-limit to the luminosity is $L_x < 5 \times 10^{34} \ erg \ s^{-1}$
($2-10$~keV, $d=4.5$~kpc). This low luminosity cannot be explained with
the high-density ($\dot{M} = 10^{-7} \ M_{\odot} \ yr^{-1}$)
wind of the circumstellar disk around the Be star. 
From optical observation during low luminosity states,
SAX~J2103.5+4545 shows $H\alpha$ in absorption,
which is interpreted as the loss of the circumstellar disk
(\cite{Reig et al. 2004}, \cite{Reig et al. 2005}).
In this case, SAX~J2103.5+4545 could accrete from the low-density wind
of the polar regions.
The mass loss rate out of circumstellar disk is $10^{-8} - 10^{-9} M_{\odot} \ yr^{-1}$,
the velocity law is $v=v_{\infty}(1-R_{Be}/r)^{\beta}$, with terminal velocity 
$v_{\infty}\approx 600 - 1800 \ km \ s^{-1}$, $\beta \approx 1$
(\cite{Waters et al. 1988}, \cite{Waters et al. 1989}). 
Assuming the SAX~J2103.5+4545 parameters found by 
Camero Arranz et al. 2007 \cite{Camero Arranz et al. 2007},
and assuming $M_{Be} = 20 \ M_{\odot}$, $R_{NS} = 10 \ km$, $M_{NS} = 1.4 \ M_{\odot}$ 
and $R_{Be}=8 \ R_{\odot}$ 
(\cite{Reig et al. 2004}, \cite{Reig et al. 2005}), 
we find that the expected luminosity 
in the Bondi-Hoyle accretion theory (\cite{Bondi and Hoyle 1944}, 
\cite{Waters et al. 1989})
is $L_x \leq 10^{34} \ erg \ s^{-1}$. 
If we consider the pulsar magnetic field of $3 \times 10^{13} \ G$,
calculated above, a wind mass loss rate 
of $8 \times 10^{-9} \ M_{\odot} \ yr^{-1}$,
and $v_{\infty}\approx 600 \ km \ s^{-1}$,
the centrifugal inibition of accretion produces an expected luminosity
$L_x = \frac{G M_{NS}}{R_m}\dot{M} \leq 10^{31} erg \ s^{-1}$,
where $R_m$ is the magnetospheric radius \cite{Davidson and Ostriker 1973}. 
These expected luminosities agree with the measured upper-limit,
and with the observed spin-down rate  
$\dot{P}_{spin-down}=(5.5 \pm 0.4) \times 10^{-9} \ s \ s^{-1}$.

\begin{acknowledgments}
LS, AP and SM acknowledge the Italian Space Agency financial
and programmatic support via contract I/008/07/0.
The autors thank the INTEGRAL Science Data Center shift team,
in particular V. Beckmann, S. Shaw and N. Produit for efficiently
sharing the Quick Look Analysis results of SAX~J2103.5+4545.
\end{acknowledgments}

\end{document}